# Comparison of the Resistivities of Nanostructured Films Made from Silver, Copper-Silver and Copper Nanoparticle and Nanowire Suspensions

Boris Polyakov[1,*], Aleksandrs Novikovs[1], Madara Leimane[1], Kevon Kadiwala[1], Martins Zubkins[1], Edgars Butanovs[1,2], Sven Oras[2,3], Elyad Damerchi[2], Veronika Zadin[2], Sergei Vlassov[3]

[1]Institute of Solid State Physics, University of Latvia, Kengaraga 8, LV-1063, Riga, Latvia

[2]Institute of Technology, University of Tartu, Nooruse 1, 50411 Tartu, Estonia

[3]Institute of Physics, University of Tartu, W. Ostwaldi 1, 50411, Tartu, Estonia

[*]corresponding author: boris@cfi.lu.lv

*corresponding author: boris@cfi.lu.lv

Abstract

Spray deposition and inkjet printing of various nanostructures are emerging complementary methods for creating conductive coatings on different substrates. In comparison to established deposition techniques like vacuum metal coating and lithography-based metallization processes, spray deposition and inkjet printing benefit from significantly simplified equipment. However, there are number of challenges related to peculiar properties and behaviour of nanostructures that require additional studies. In present work, we investigate electroconductive properties and sintering behaviour of thin films produced from nanostructures of different metals (Ag, Cu and Cu-Ag) and different shapes (nanowires and spherical nanoparticles), and compare them to the reference Ag and Cu magnetron deposited films. Synthesized nanostructures were studied with transmission electron microscopy. Morphology and crystallinity of produced metal films were studied with scanning electron microscopy and X-ray diffraction. The electrical parameters were measured by the van der Pauw method. All nanowires-based films provided high conductivity and required only modest thermal treatment (200 ºC). To achieve sufficient sintering and conductivity of nanoparticles-based films, higher temperatures are required (300 ºC for Ag nanoparticles and 350 ºC for Cu and Cu-Ag nanoparticles). Additionally, stability of nanowires was studied by annealing the samples in vacuum conditions inside a scanning electron microscope at 500 ºC.

**Introduction**

Spray deposition and inkjet printing of various nanostructures are emerging complementary fabrication methods of additive manufacturing for creating functional coatings [1,2]. Digital inkjet printing allows precise jetting of uniform droplets onto different substrates such as flexible plastic sheets, glass, silicon, etc [3,4]. Small droplets (down to 1 picoliter) are produced by small diameter nozzles, which place strong requirements on both nanostructure size and physical properties of inks (viscosity, surface tension, boiling temperature) [4]. Spray deposition of functional inks has compromised resolution compared to inkjet printing, but can be used for covering of large area substrates, and benefits from less strict requirements for nanomaterials size. In particular, it allows to use nanowires (NWs) up to few tens of µm long, which is impossible for inkjet printing. Both techniques can be used for creating electrically conductive coatings by using metal nanostructures for deposition. Compared to well-established vacuum metal coating and lithography-based metallization processes [1,5], where metals are deposited from bulk metal targets using complicated manufacturing equipment (magnetrons, e-beam and thermal evaporators), spray deposition and printing benefit from significantly simpler equipment but are more demanding in choosing the materials and often involve post processing like sintering or annealing to improve electrical conductivity and performance of the coating [6–9].

When speaking about thermal treatment of deposited or printed nanoparticles (NPs) and NWs, it has to be considered that behaviour of metal nanostructures under elevated temperatures differs significantly from that of their bulk counterparts being not only material, but also size and shape dependant [10,11]. In general, metal nanostructures melt at lower temperatures compared to bulk, and the melting temperature decreases with reduction of the size and dimensionality (NPs in general have lower melting temperature than NWs of the same diameter) of the structures [10,12,13]. Moreover, in case of prolonged heat treatment (minutes and above) heat-induced diffusion of surface atoms can result in morphological changes and sintering of nanostructures at temperatures far below melting point of the material [14,15]. In contrast to situation with melting, NWs are more subject to heat-induced diffusion than NPs. Sintering of NWs at intersections [16], which is beneficial for improving electrical conductivity, can happen at very mild heating (e.g., as low as 125 – 200 ºC for Au and Ag NWs [15]). However, further increase in temperature and/or heating time can result in splitting of NWs into shorter fragments driven by Rayleigh instability and energy minimisation via spheroidization, which will lead to decrease in electrical conductivity [16,17].

In present work we investigate electroconductive properties and sintering behaviour of metal thin films produced from nanostructures of different metals (Ag, Cu and Cu-Ag) and different shapes (NWs and spherical NPs). In majority of reported works researchers are focused on one type of nanostructures (either NPs or NWs), or only one metal, while our goal was to measure electrical properties of both NWs and NPs using the same characterization equipment and substrate to find out advantages and disadvantages of each metal material or nanostructure type from the application point of view. Cu and Ag nanostructures were chosen as most popular materials for conductive inks, and as example of oxygen sensitive and insensitive nanomaterials respectively. Ag plating of Cu nanostructures is frequently used to increase oxidation stability of Cu nanostructures [18,19]. Results obtained on nanostructures are compared to the reference Ag and Cu magnetron deposited films.

**Experimental**

Suspension of Ag NWs (nominal diameter 35 nm) in ethanol was purchased from Blue Nano Inc. All other chemicals were purchased from Sigma-Aldrich: copper chloride dehydrate ($CuCl_2·2H_2O$) (≥99.0%), hexadecylamine (98%), glucose (≥99.5%), oleylamine (technical grade, 70%), silver nitrate ($AgNO_3$) (≥99.0%), monoethanolamine (≥98%), poly(acrylic acid) (PAA, Mw 3000), ethanol (absolute), copper acetylacetonate ($Cu(acac)_2$) (97%), acetone (≥99.9 %), hexane (anhydrous, 95%), toluene (anhydrous, 99.8%).

Cu NWs were synthesized using hydrothermal method similar to procedure described in [19,20] with small modifications. Namely, precursor solution was prepared by dissolving of 2.72 g $CuCl_2·2H_2O$, 0.28 g glucose and 1.8 g hexadecylamine (HDA) in 100 ml deionized water. After continuous stirring for 2 h at room temperature, the solution gradually turned into blue emulsion. Then, the precursor solution was transferred into a 200 ml Teflon-lined autoclave and heated for 24 h at 120 °C. After the autoclave was cooled to room temperature, the reddish reaction product was washed three times with water and toluene respectively. Finally, the Cu NWs were re-dispersed in toluene for further use.

Cu-Ag NWs were prepared by modification of method described in [18], which previously was applied for Cu NPs. To prepare silver plating solution, 0.018 g of $AgNO_3$ was dissolved in 5 mL of oleylamine, and then added to the flask containing Cu nanowire toluene suspension, which was then heated to 80 °C for 2 h. Then, the Cu–Ag core–shell NWs were centrifuged at 6 000 rpm for 10 min and washed three times with toluene. Note, that obtained Cu-Ag NWs have much smoother surface in comparison to other methods (e.g. [17,19]). Obtained Cu-Ag NWs were redispersed in toluene.

Ag NPs were synthesized by using the procedure described in [21]. Firstly, monoethanolamine (MEA, 25.00 g) was dissolved in 60 ml of deionized water to make a homogeneous clear solution, after that polyacrylic acid (PAA, 3.00 g) was added to solution and was stirred to make a homogeneous solution. Then silver nitrate ($AgNO_3$, 17.00 g) was dissolved in this solution, and stirred for 1 h at room temperature to obtain yellowish and transparent solution. The resultant solution was then gradually heated to 85°C with continuous stirring for 1h, solution become dark brown. The solutions were precipitated by the addition of ethanol (EtOH) and stirred for 1-2 min. The black precipitates were formed and collected through centrifugation were once washed (6000 rpm, 30 min) with DI $H_2O$ and ethanol (DI $H_2O$/ EtOH) and then four times just with ethanol (EtOH). Then, the black precipitates were dispersed in 10 mL DI $H_2O$ and left for drying at 75°C for 24 h. Finally, grey solid Ag precipitates (powder) were obtained. Before use Ag nanoparticles were re-dispersed in ethanol.

Cu and Cu-Ag NPs were synthesized by the thermal decomposition method and following Ag shell overgrowth as described in [18]. Precursor solution prepared from 0.83 g of $Cu(acac)_2$ dissolved in 50 mL of oleylamine in a 100 mL three-neck flask. The solution was kept under N2 gas flow and constant stirring at 100 °C for 30 min until complete dissolution of $Cu(acac)_2$. Then, the temperature was increased to 235 °C and held at this temperature for 3 h under $N_2$ gas protection. The solution become metallic-red, indication formation of Cu nanoparticles. Then, the flask was cooled down to room temperature. Then, the Cu-containing solution was divided into two portions. The first portion was diluted by same amount of hexane and centrifuged 10 min at 6000 rpm, and re-dispersed in toluene. In contrast to reference [18], where only centrifugation was used for sedimentation of Cu and Cu-Ag nanoparticles, we used acetone – toluene washing approach. Acetone was added 1:1, and centrifuged 10 min at 6000 rpm, and re-dispersed in toluene again (repeated twice). The second portion was used for preparation of Cu-Ag nanoparticles. Silver plating solution prepared by dissolving 65 mg of $AgNO_3$ in 19 ml of oleylamine and added to the flask with Cu nanoparticles. The flask was heated to at 80 °C and held at this temperature for 2 h. Colour of solution become reddish-brown. Then, the resulting solution was diluted by hexane, centrifuged and re-dispersed in toluene. Washing procedure identical to Cu NPs. Final product re-dispersed in toluene.

For electrical characterization, X-ray diffraction (XRD) and scanning electron microscopy (SEM) measurements, NWs or NPs were drop casted from suspension onto 1x1 cm $Si/SiO_2$ (Semiconductor wafers LTD) wafers. Three samples of each type were prepared to increase reliability of electrical measurements. Samples were then annealed in order to improve electrical contact. Ag NWs and Ag NPs were annealed in atmosphere for 30 min at different temperatures ranging from 200 to 350 °C. Sintering of Cu and Cu-Ag nanostructures was carried in argon atmosphere inside glovebox to avoid Cu oxidation [18]: NWs were annealed for 30 min and NPs for 2 h 30 min. Thickness determination for films made from NPs and NWs was based on weighing of samples before and after deposition and annealing of nanostructures, following by calculation of apparent bulk equivalent film thickness based on measured mass and known material density and sample size. Mass of deposited metal nanomaterials was determined by weighting the substrates before ($m_1$) deposition and then after ($m_2$) deposition and sintering using precision scales (A&D Company Limited, HR-150A). The mass difference was used for apparent thickness ($d$) determination using material density [22] and sample size (1x1cm, $a$=1cm): $\rho=m/V$, $V=a*a*d$, $m=m_2-m_1$.

Silver content in Cu-Ag NPs and NWs was measured by Energy Dispersive X-ray (EDX) microanalysis. Copper density: 8.96 g $cm^{-3}$. Silver density: 10.5 g $cm^{-3}$. Copper/silver NPs (64%/26%): 9.3 g $cm^{-3}$. Copper/silver NWs (88%/12%): 9.14 g

cm$^{-3}$. No sputter cleaning of the samples was performed prior to analysis. The measured spectra were calibrated relative to the adventitious C 1s peak at 284.8 eV.

For transmission electron microscopy (TEM) measurements, diluted suspensions of metal nanostructures were drop-casted on carbon-coated copper grids (Lacey Carbon 200 Mesh Cu, Agar Scientific).

To observe the dynamics of heat-induced morphological changes in investigated nanostructures in vacuum conditions heating experiments were performed also inside SEM (FEI Nanosem 450, US) using dedicated SEM heating stage (Kammrath-Weiss, Germany) allowing heating up to 800 °C.

For reference electrical measurements, Ag and Cu thin films were deposited by magnetron sputtering (from Ag and Cu targets respectively) on 1x1 cm$^2$ Si/SiO$_2$ wafers. Magnetron sputtering was performed in the vacuum system SAF25/50 (Sidrabe) at 100 W DC magnetron power in a mixed atmosphere of sputtering Ar gas (30 sccm) at a total pressure of 2.7 x 10$^3$ Pa. The substrate was mounted 15 cm above the magnetrons and maintained at room temperature during the deposition process. The film thickness was measured by Dektak 150 profilometer (Veeco).

The structure and crystallinity of the films were determined by the X-ray diffraction (XRD) technique. The XRD patterns were recorded using a benchtop Rigaku MiniFlex 600 powder diffractometer in Bragg-Brentano θ-2θ geometry and 600 W Cu anode (Cu Kα radiation, λ = 1.5406 Å) X-ray tube. The film morphology and element composition were studied with a SEM/FIB (Lyra XM, Tescan) at 12 kV equipped with Energy-dispersive X-ray (EDX) detector (EDX Oxford X-Max 50 mm$^2$ detector). The crystalline structure of the films was imaged using TEM (Tecnai GF20, FEI) operated at a 200 kV accelerating voltage. The electrical parameters were measured in the van der Pauw configuration using a Hall effect system, HMS5000 (Ecopia).

The chemical composition of the NW samples was confirmed with X-ray photoelectron spectroscopy (XPS) measurements performed using an X-ray photoelectron spectrometer ESCALAB Xi (ThermoFisher). Al Kα X-ray tube with the energy of 1486 eV was used as an excitation source, the size of the analysed sample area was 650 μm x 100 μm and the angle between the analyser and the sample surface

Thin Solid Films, Volume 784, 1 November 2023, 140087, https://doi.org/10.1016/j.tsf.2023.140087was 90°. An electron gun was used to perform charge compensation. The base pressure during the spectra acquisition was better than $10^{-5}$ Pa.

**Results and Discussion**

The phase composition of the as-prepared nanostructures deposited on $SiO_2/Si$ substrates before and after annealing was confirmed using XRD. In the case of Ag NWs and NPs (Figure 1a,b), the two Bragg peaks in the diffraction pattern at 37.9 and 44.3 degrees were attributed to silver (111) and (200) planes, respectively (ICDD-PDF #04–0783). Similarly, for Cu NWs and NPs (Figure 1c,d) the two peaks centred at 43.3 and 50.4 degrees belong to (111) and (200) planes (ICDD-PDF #04–0836). The broad lower intensity shoulder for the Cu (200) reflection in the NW pattern could be related to the NPs that were a side-products of the synthesis procedure. Furthermore, a broad peak (marked by asterisk at Figure 1d) can be seen around 37.5 degrees in the Cu NPs pattern, which could indicate formation of a CuO phase (undesirable oxidation of the metal NPs). As for Cu-Ag core-shell nanostructures (see Figure 1e,f), Bragg peaks from both phases were identified. Both Cu (111) and (200) peaks are present in the patterns, while only Ag (111) peak with a lower intensity can be clearly distinguished (the Ag (200) peak overlaps with the stronger Cu (111) peak). While no distinct change in XRD patterns before and after annealing can be observed for NWs, a significant narrowing of Ag and Cu-Ag NPs XRD peaks confirm successful sintering of the NPs, and the modest narrowing of Cu NPs XRD peaks reflects the poor sintering of NPs. XRD patterns for the magnetron-deposited Ag and Cu films give qualitatively similar results. Two peaks for (111) and (200) are present on XRD patterns of both Ag and Cu films. The Bragg peaks at around 33 degrees present in all patterns were attributed to the Si (100) substrate (forbidden Si (200) reflection) [23].

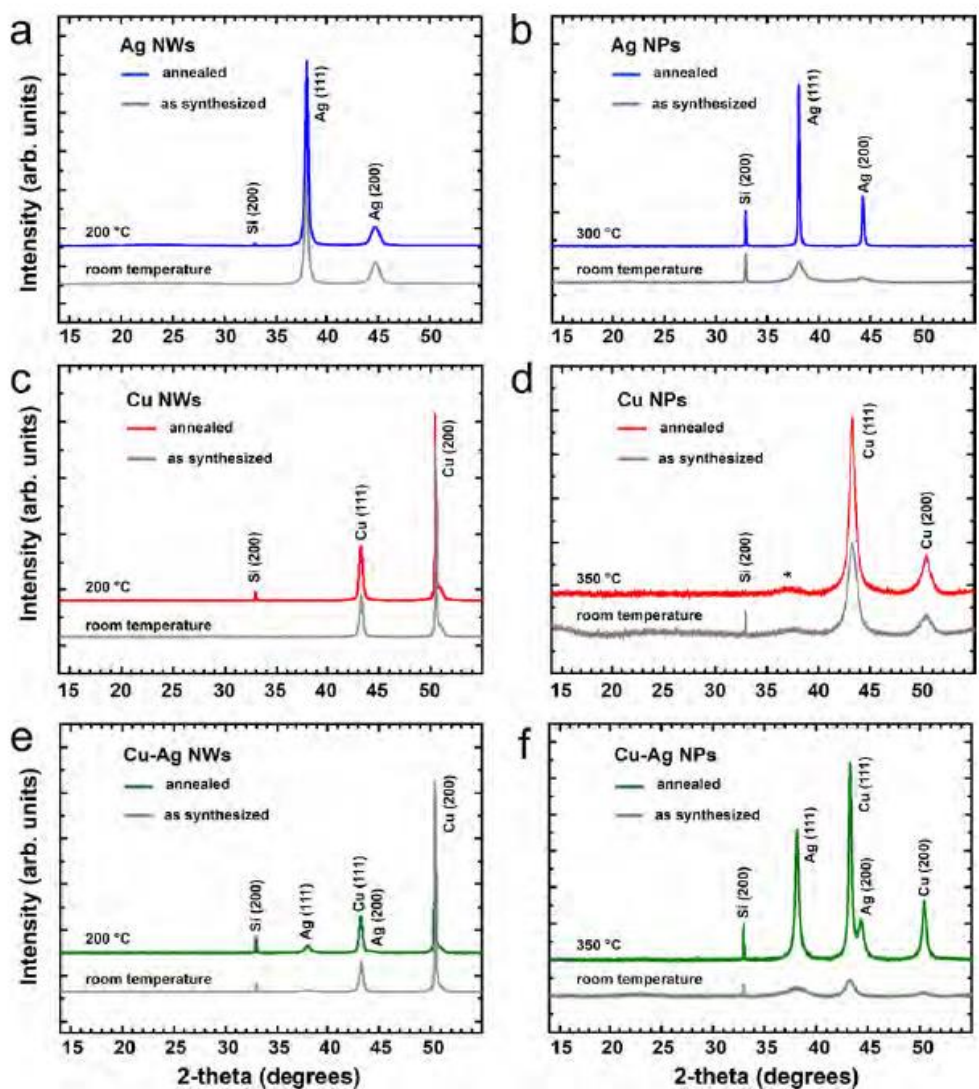

*Figure 1.* XRD patterns of (a,b) Ag, (c, d) Cu and (e, f) Cu-Ag NWs and NPs on SiO$_2$/Si substrates before and after annealing.

In addition to XRD measurements, XPS analysis was performed in order to verify the chemical states of the metal elements in the nanostructures (Figure 2). High-resolution spectra of Cu 2p and Ag 3d were acquired and calibrated relative to the adventitious C 1s peak at 284.8 eV. Qualitatively similar results were obtained for NPs and NWs. In Cu and Cu-Ag nanostructures Cu 2p$_{3/2}$ peak was located at 932.8-932.9 eV (spin-orbit splitting $\Delta_{3/2-1/2}$ = 19.75 eV), matching metal Cu states, and no distinctive satellite features can be observed around 945 eV, indicating that oxide states are not present. As for Ag 3d scans in Ag and Cu-Ag nanostructures, Ag 3d$_{5/2}$ peak was located at 368.2-368.3 eV (spin-orbit splitting $\Delta_{5/2-3/2}$ = 6.0 eV), which corresponds to Ag metal state. No sulfurization of Ag was observed. The O 1s scans only indicated significant contributions from organic surface contaminants and Si/SiO$_2$ substrate, thus not giving valuable information on metal nanostructure chemistry.

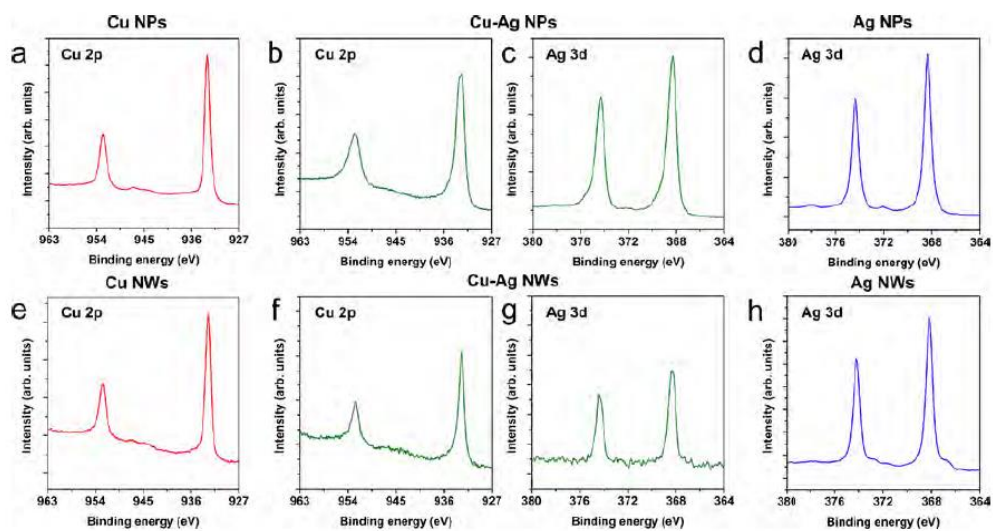

*Figure 2. High-resolution XPS spectra of constituent elements for (a) Cu, (b,c) Cu-Ag, (d) Ag NPs, and (e) Cu, (f,g) Cu-Ag, (h) Ag NWs.*

Average diameters of NWs as found from TEM images are following: Ag NWs – 27.2 nm, Cu NWs – 47.2 nm, Cu-Ag NWs – 42.5 nm. The surfaces of Ag and Cu NWs were very smooth, while surface of Cu-Ag NWs was rougher due to Ag plating procedure (see Figure 3a,b,c for chosen TEM images). According to SEM images, the length of Ag, Cu and Cu-Ag NWs exceeded 10 μm. After 30 min of annealing at 200ºC, Cu and Cu-Ag NWs were intact, while Ag NWs started to diffuse at intersections with other Ag NWs and formed thicker segments (see Figure 3d,e,f). According to EDX measurements performed in SEM, Ag content in Cu-Ag NWs is around 12%.

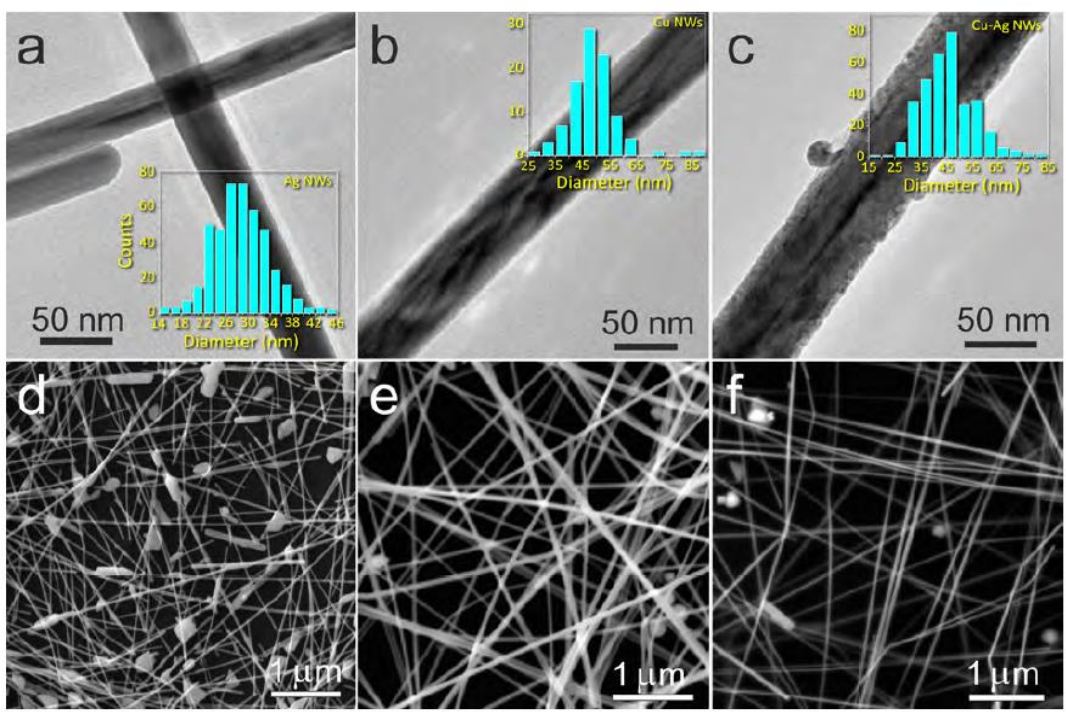

*Figure 3. Chosen SEM and TEM images of (a,d) Ag, (b,e) Cu and (c,f) Cu-Ag NWs annealed at 200 ºC. Corresponding diameter distribution histograms are shown as inset in SEM images.*

Average diameters of NPs as found from TEM images are following: Au NPs – 10.8 nm, Cu NPs – 11.1 nm, Cu-Ag NPs – 12.5 (see Figure 4a,b,c for chosen TEM images). According to SEM images of samples with NPs after heat treatment (Figure 4d,e,f), Ag NPs started to sinter at 200 ºC, and formed well-sintered film at 300 ºC (Figure 4a,d). Cu NPs did not sinter completely even at 350 ºC and separate particles are still distinguishable (Figure 4h) while Ag-coated Cu NPs sinter well at 350 ºC (Figure 4f,i). According to EDX measurements, Ag content in Cu-Ag film is around 26%.

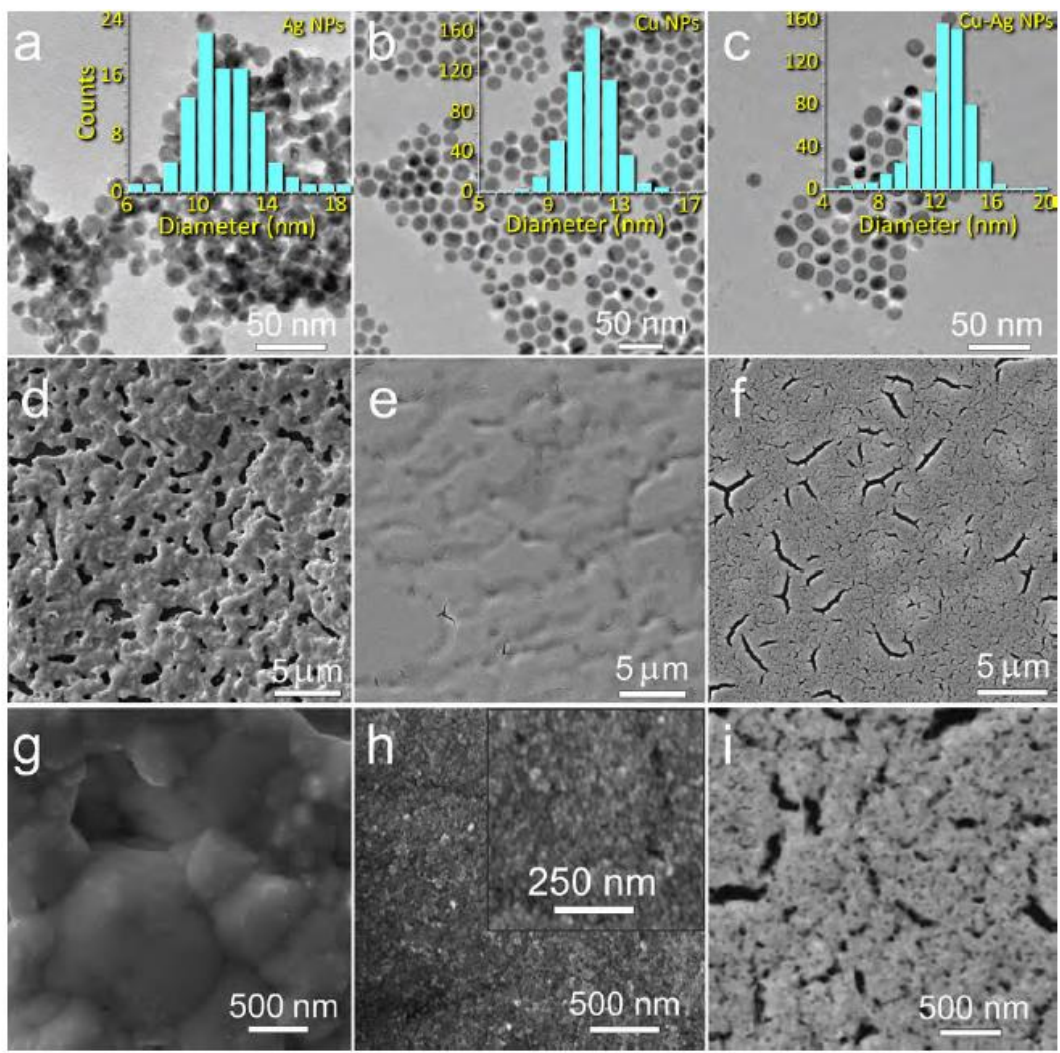

*Figure 4.* TEM (a-c) and SEM (d-i) images of (a,d,g) Ag, (b,e,h) Cu, and (c,f,i) Cu-Ag NPs with corresponding histograms of NPs size distribution. TEM images are from particles without thermal treatment. SEM images are taken after sintering.

Next, conductivities of inks made of different metal nanomaterials were measured and compared with those obtained on magnetron deposited metal thin films. Results on resistance, resistivity and film thickness of all measured types of investigated metal films are summarized in Table 1. The thickness of magnetron deposited Ag and Cu films was measured directly by surface profilometer. Thickness determination of films produced from nanomaterials (NWs and NPs) is complicated due to high porosity of such films. We applied an indirect method of apparent thickness determination, which was based on weighing of samples before and after deposition and annealing of nanostructures, following by calculation of apparent film thickness based on measured mass and known material density and sample size.

Resistivity of magnetron-deposited Ag and Cu thin films were $2.6 \times 10^{-6}$ Ω·cm and $5.9 \times 10^{-6}$ Ω·cm respectively, which agrees well with literature data for magnetron deposited Ag thin film ($2.0 \times 10^{-6}$ Ω·cm) [24] and magnetron deposited Cu thin film ($5.2 \times 10^{-6}$ Ω·cm) [25].

As-deposited Ag NWs film have sheet resistance $7.3 \times 10^{-1}$ Ω/sq (corresponding $2.2 \times 10^{-5}$ Ω·cm for calculated thickness 300 nm). After 30 min annealing at 200 ºC resistance slightly increased to 1.1 Ω/sq (corresponding 3.3-5 Ω·cm) due to fragmentation of some NWs (Figure 3a). This process is driven by surface atom diffusion and it is a known phenomenon [15], therefore it is important to find optimal balance between just sintering and fragmentation, which was outside the scope of the present paper. Electrical measurements of as-deposited Cu NWs film had unstable results and were not included in the Table 1. After 30 min annealing at 200 ºC in inert (Ar) atmosphere the resistance of Cu NW film was 1.1 Ω/sq (corresponding $1.2 \times 10^{-5}$ Ω·cm for calculated thickness 110 nm). As deposited Cu-Ag NWs film had resistance $7.9 \times 10^{-1}$ Ω/sq (corresponding to $3.9 \times 10^{-5}$ Ω·cm for calculated thickness of 495 nm). After 30 min of annealing at 200 ºC in Ar atmosphere the resistance of Cu-Ag NWs film slightly decreased to $6.5 \times 10^{-1}$ Ω/sq (corresponding $3.2 \times 10^{-5}$ Ω·cm).

As-deposited Ag NPs film have high sheet resistance $5.4 \times 10^{5}$ Ω/sq (corresponding 8.1 Ω·cm for calculated thickness 150 nm). After 30 min of annealing at 200 ºC in atmosphere the resistance of Ag NPs film decreased significantly to 57 Ω/sq (corresponding to $8.5 \times 10^{-4}$ Ω·cm). Ag NPs sintering is clearly visible at the SEM image (Figure 4a). As-deposited Cu and Cu-Ag NPs films were not conductive. After 150 min of annealing at 350 ºC in Ar atmosphere resistance of Cu and Cu-Ag NPs films dropped to $1.41 \times 10^{2}$ (corresponding to $1.9 \times 10^{-2}$ Ω·cm for 1344 nm) and 1.1 Ω/sq respectively (2.6E-4 Ω·cm for 1760 nm). High resistivity of Cu NPs based films can be explained by poor sintering of NPs (Figure 4e), probably due to surface oxidation. Silver coating of Cu NPs allows to avoid surface oxidation of NPs and achieve much better sintering and lower resistivity consequently (Figure 4f).

It should be noted that conductivity in nanostructured metal films is dictated by carrier mobility rather than by carrier concentration, and mainly affected by electron scattering at grain boundaries or nanostructures surface. From this point of view, long metal NW networks are excellent conductive electrodes; conductivity of metal NPs films strongly affected by sintering degree.

Table 1. Comparison of resistance and resistivity for different metal films.

| Sample | Thickness (nm) | Before heating | | After heating | | Anneal.temp., ºC | Time, min |
|---|---|---|---|---|---|---|---|
| | | Resistivity (Ω·cm) | Sheet resistance (Ω/sq) | Resistivity (Ω·cm) | Sheet resistance (Ω/sq) | | |
| Ag magnetr. | 91 | $2.6 \times 10^{-6}$ | $2.9 \times 10^{-1}$ | - | - | - | - |
| Cu magnetr. | 112 | $5.9 \times 10^{-6}$ | $5.3 \times 10^{-1}$ | - | - | - | - |
| Ag NWs | 300* | $2.2 \times 10^{-5}$ | $7.3 \times 10^{-1}$ | $3.3 \times 10^{-5}$ | 1.1 | 200 | 30 |
| Cu NWs | 110* | - | - | $1.2 \times 10^{-5}$ | 1.1 | 200 | 30 |
| Cu-Ag NWs | 495* | $3.9 \times 10^{-5}$ | $7.9 \times 10^{-1}$ | $3.2 \times 10^{-5}$ | 0.65 | 200 | 30 |
| Ag NPs | 150* | 8.1 | $5.4 \times 10^{5}$ | $8.5 \times 10^{-4}$ | 57 | 300 | 30 |
| Cu NPs | 1344* | - | - | $1.9 \times 10^{-2}$ | 141 | 350 | 150 |
| Cu-Ag NPs | 1760* | - | - | $2.6 \times 10^{-4}$ | 1.1 | 350 | 150 |

*Thickness calculated from metal mass measurements.

Finally, we tested morphological stability of Ag, Cu and Cu-Ag NWs in real time during 30 min annealing at 500 ºC inside SEM (Figure 5) to get deeper insight into the dynamics of the diffusion processes. According to observed results, both Ag and Cu-Ag NWs start to melt at intersection areas, however uncoated Cu NWs remain almost intact (at close observation the intersected NWs are welded together). It may be related to the fact that uncoated Cu NWs have native oxide layer that inhibits the surface diffusion of Cu atoms, while Ag plating prevents formation of oxide and may even promote diffusion and fragmentation due to inner stresses caused by crystal lattice mismatch between Cu core and Ag shell. In general, we would like to emphasise that observed heat-induced morphological changes in NWs (fragmentation, thickening, etc) should not be confused with melting but should be attributed to diffusion of atoms. By receiving the additional energy in the form of heat, atoms in NWs gain the ability to overcome diffusion barriers and rearrange the structure into shorter and more rounded fragments driven by Rayleigh instability phenomena and energy minimization principle. These processes discussed in more details e.g. in Vigonski et. al [15].

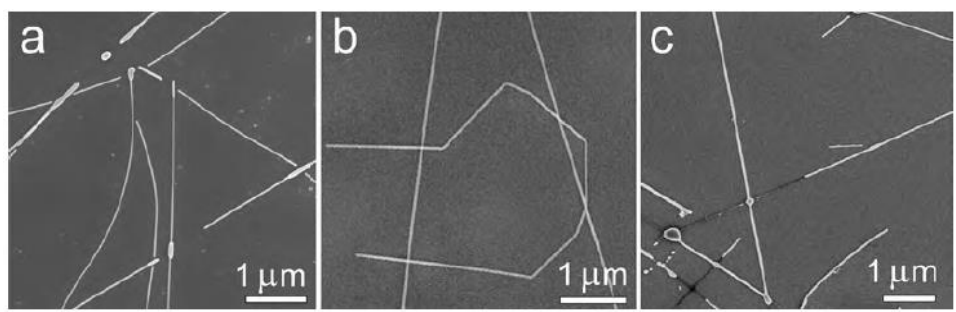

*Figure 5.* SEM images of (a) Ag, (b) Cu and (c) Cu-Ag NWs annealed at 500 ºC.

We also compared the behaviour of magnetron-deposited Ag and Cu thin films in 30 min annealing at 450-500 ºC in inert atmosphere. For both films dewetting was observed, resulting in transformation of uniform film into large separated grains (Figure 6). It suggests that at least for Cu, films deposited from NWs have higher thermal stability than magnetron-deposited films. Note that the film stability depends not only on the annealing temperature, but also on the film thickness: compared magnetron deposited Cu film was 112 nm thick, and Cu NWs film has apparent thickness 110 nm; for films of different thickness stability results can differ [26].

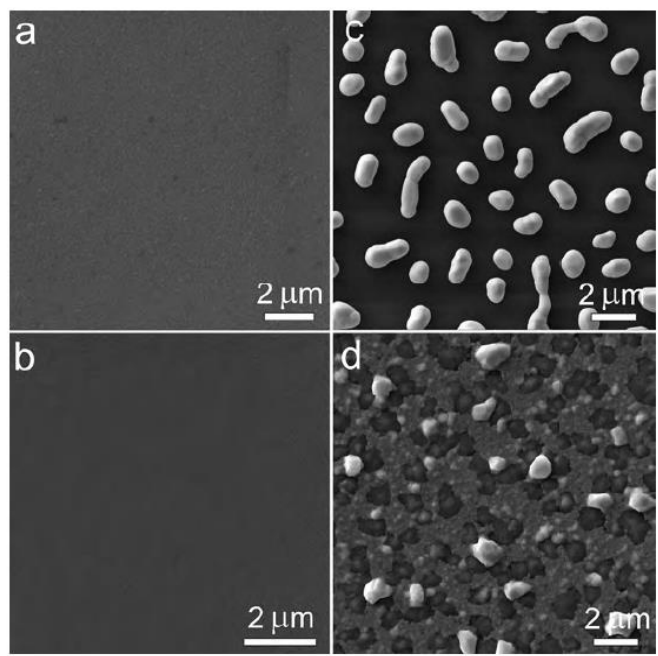

*Figure 6. SEM images of Ag and Cu thin film deposited by magnetron before annealing (a) and (b) respectively; after annealing (c) and (d).*

**Conclusions**

In this work, we compared resistivity of thin films produced from Ag, Cu, and Cu-Ag NWs, and Ag, Cu, and Cu-Ag NPs with resistivity of Ag and Cu thin films deposited by magnetron sputtering. Resistivity of Ag, Cu and Ag-Cu NWs films ($2.2 \times 10^{-5}$, $1.2 \times 10^{-5}$ and $3.2 \times 10^{-5}$ $\Omega\cdot$cm respectively) was approximately order of magnitude higher in comparison to magnetron deposited Ag and Cu thin films ($2.6 \times 10^{-6}$ and $5.9 \times 10^{-6}$ $\Omega\cdot$cm respectively). While resistivity values of Ag, Cu and Ag-Cu NPs based films were significantly higher ($8.5 \times 10^{-4}$, $1.9 \times 10^{-2}$ and $2.6 \times 10^{-4}$ $\Omega\cdot$cm respectively). Moreover, Ag NWs and Cu-Ag NWs films can be used without sintering. Interesting to note, that Ag NWs films start to degrade at 30 min annealing at 200 ºC, while no significant morphological changes were detected in Cu and Cu-Ag NWs films at this temperature. Consequently, resistivity of Ag NWs film slightly increases upon annealing, while resistivity of Cu and Cu-Ag NWs films decreases due to evaporation of organic surfactant and welding NWs together at the cross-points. Moreover, Cu NWs stay stable even at 500 ºC, in contrast to magnetron deposited Cu thin film, making Cu NWs an attractive alternative to Cu films in applications that involve elevated temperatures. Ag NPs based film requires much shorter sintering time (30 min at 300 ºC), while Cu and Cu-Ag NPs sintering was 5 times longer (150 min at 350 ºC). No full sintering was achieved for Cu NPs, probably due to surface oxidation. Important to note, that sintering of both Cu and Cu-Ag NWs and NPs films require inert atmosphere to avoid oxidation. Summarising the above-mentioned results, we can conclude that Ag NPs provide both good conductivity and low sintering temperature, while Ag NWs also provide excellent conductivity, but unable to withstand heat treatment. Cu and Cu-Ag NWs both able to provide good conductivity and have low sintering temperature.

**Acknowledgment**

This research is funded by the ERDF project "Functional ink-jet printing of wireless energy systems" No. 1.1.1.1/20/A/060. S.V, S.O., E.D. and E.B. was supported by the European Union's Horizon 2020 program, under Grant Agreement No. 856705 (ERA Chair "MATTER"). Institute of Solid State Physics, University of Latvia as the Center of Excellence has received funding from the European Union's Horizon 2020 Framework Programme H2020-WIDESPREAD-01-2016-2017-TeamingPhase2 under grant agreement No. 739508, project CAMART2. The authors are grateful to Annamarija Trausa and Luize Dipane for help with sample preparation and characterization.